\documentclass[11pt,a4paper]{article}
\usepackage{amsmath}

\usepackage[psamsfonts]{amssymb}
\usepackage{amsmath}
\usepackage{epsfig}

\author{H. Mohseni Sadjadi\footnote{mohsenisad@ut.ac.ir}
\\ {\small Department of Physics, University of Tehran,}
\\ {\small P. O. B. 14395-547, Tehran 14399-55961, Iran}}
\title{Crossing the Phantom divide line in the Chaplygin gas model}

\begin{document}
\maketitle
\begin{abstract}
The r\^{o}le of the interaction in reaching and crossing the
phantom divide line in the Chaplygin gas model is discussed. We
obtain some necessary properties of the interaction that allow the
model to arrive at or cross the phantom divide line. We show that
these properties put some conditions on the ratio of dark matter
to dark energy density in the present epoch.

\end{abstract}

\section{Introduction and preliminaries}
To describe the present accelerated expansion of the universe
\cite{acc}, many models have been introduced. In one of these
pictures nearly 70\% of the universe is filled by an exotic smooth
energy component with negative pressure dubbed as dark energy. One
of the candidates for the dark energy is the Chaplygin gas
\cite{chap}, a perfect fluid whose equation of state (EoS) is
given by
\begin{equation}\label{1}
P_d=-{A\over \rho_d}.
\end{equation}
$A$ is a positive real constant and $P_d$ and $\rho_d$ are the
pressure and the energy density of the dark energy respectively in
the rest frame of the fluid. The Chaplygin gas plays a dual
r\^{o}le: it behaves as a dustlike matter in the early era (i.e.
for small scale factor $a$), and as a cosmological constant at
late times (i.e. for large values of $a$). However, this unified
model of dark energy and dark matter, suffers from problems such
as production of oscillations or exponential blowup of the matter
power spectrum which is inconsistent with observation \cite{mass}.
Besides this noninteracting model is unable to describe the
coincidence problem \cite{coin}, i.e. this model cannot explain
why the densities of the dark matter and dark energy are of the
same order today. This lies on the fact that energy density of the
non interacting cold dark matter (whose EoS parameter is
$w_m\simeq 0$) redshifts faster than the energy density of the
dark energy sector (whose EoS parameter satisfies $w_d<-1/3$), so
the ratio of matter to the dark energy density is expected to
decrease rapidly as the universe expands in noninteracting models.
In addition this model is inconsistent with the phantom divide
line crossing which based on astrophysical data seems to be
occurred in the present era \cite{cross}.

In this paper we assume that the universe is a spatially flat
Friedmann-Robertson-Walker space-time which besides the dark
energy component, whose EoS is given by (\ref{1}), is composed of
the cold dark matter interacting with the Chaplygin gas (which is
taken as the dark energy sector) via the interaction term $Q$:
\begin{eqnarray}\label{3}
&&\dot{\rho_d}+3H(-{A\over \rho_d^2}+1)\rho_d=-Q\nonumber \\
&&\dot{\rho_m}+3H\rho_m=Q.
\end{eqnarray}
$\rho_m$ is the density of the pressureless (cold) dark matter.
$H$ is the Hubble parameter and "dot" denotes the derivative with
respect to the comoving time. The presence of the interaction
source, by permitting the energy exchange between dark matter and
dark energy, may alleviate the coincidence problem \cite{amen},
meanwhile, as we will see, depending on the form of interaction,
may allow the phantom divide line crossing.

The Friedmann equations are
\begin{eqnarray}\label{4}
\dot{H}&=&-4\pi({-A\over \rho_d}+\rho_d+\rho_m)\nonumber \\
H^2&=&{8\pi\over 3}(\rho_d+\rho_m).
\end{eqnarray}
The condition $\rho_m+\rho_d=\rho_c:={3H^2\over {8\pi}}$
guarantees the flatness of the universe.   The universe is
accelerating if
\begin{equation}\label{5}
0<\rho_d<{-\rho_m+\sqrt{\rho_m^2+12A}\over 2}.
\end{equation}
In the absence of the interaction, i.e. $Q=0$, $\rho_d$ is given
by
\begin{equation}\label{6}
\rho_d=\sqrt{A+Ca^{-6}},
\end{equation}
where $C$ is a real positive constant. The (EoS) parameter of dark
energy is
\begin{equation}\label{2}
w_d:={P_d\over \rho_d}=-{A\over \rho_d^2}.
\end{equation}
For (\ref{6}) we have $-1<w_d<0$ for $a\in(0,\infty)$, hence
$w_d=-1$ cannot be crossed in this model. As a consequence the EoS
parameter of the universe $w=\Omega_d w_d$, where
$\Omega_d={\rho_d\over \rho_c}<1$, cannot cross the phantom divide
line. Using $\lim_{a\to \infty}\rho_d=\sqrt{A}$, one can show that
the ratio of dark matter to dark energy density, defined by
\begin{equation}\label{7}
r={\rho_m \over \rho_d},
\end{equation}
becomes negligible at late time: $lim_{a\to \infty}r=0$. This was
expected because, as was mentioned, cold dark matter redshifts
faster than dark energy component. Now to investigate what happens
in the presence of the interaction \cite{inter}, let us consider
the simple interaction term
\begin{equation}\label{8}
Q=3\Gamma H\rho_d,\,\,\,\, \Gamma>0.
\end{equation}
In this model the solution to the first equation in (\ref{3}) is
\begin{equation}\label{9}
\rho_d=\sqrt{{A\over {1+\Gamma}}+Ca^{-6(\Gamma+1)}},
\end{equation}
where $C$ is a real positive constant. In this case in contrast to
the noninteracting case where $w_d$ was restricted to the set
$(-1,0)$, in principle the EoS parameter of the Chaplygin gas can
cross the phantom divide line ($w_d<-1$), for
\begin{equation}
a>\left({\Gamma A\over C(1+\Gamma)} \right)^{-1\over 6(\Gamma+1)}.
\end{equation}

Obtaining an exact solution for $\rho_m$ or $H$ is not
straightforward, but asymptotic solutions can be derived. Using
\begin{equation}\label{12}
{dH^2\over dx}=-8\pi(-{A\over \rho_d}+{3H^2\over 8\pi}),
\end{equation}
where $x:=\ln a$, we arrive at $\lim_{a\to \infty}
\rho_c=\sqrt{(\Gamma+1)A}$. In this limit the ratio of dark matter
to dark energy, $r:={\rho_m\over \rho_d}={\rho_c\over \rho_d}-1$,
obeys $\lim_{a\to \infty}r=\Gamma$. In addition the EoS parameter
of the universe satisfies $\lim_{a\to \infty}w=-1$. In
\cite{stable}, where the r\^{o}le of the interaction on dynamical
evolution of Chaplygin gas was studied in details, it was shown
that this is a stable critical point corresponding to a stable
attracting scaling solution in which although the Chaplygin gas
crosses the phantom divide line but the universe will enter to a
de Sitter phase and the big rip is avoided.

So, the presence of the interaction term, which was first
introduced to alleviate the coincidence problem, may allow the
dark energy to cross the phantom divide line. In the next section
we will find some {\it necessary} conditions for $Q$ to allow the
dark energy to reach or cross the phantom divide line and using an
example show that there may be a deep relationship between the
coincidence problem and $w_d \simeq -1$ in the present epoch.

\section{The r\^{o}le of the interaction in crossing the phantom divide line and the
coincidence problem}

In the presence of an arbitrary interaction, obtaining an exact
solution to the Friedmann equations (to see whether the model can
reach or cross $w_d=-1$), may not be generally possible. In this
part, instead of solving these equations, we try to obtain some
{\it necessary} conditions which must be satisfied if the model
reach $w_d=-1$. These conditions put some constraints on $Q$ and
the parameters of the model. Besides, depending of the form of
$Q$, $w_d$ may be a decreasing function at $t=t_0$, where $t_0$ is
defined by $w_d(t_0)=-1$. In this situation $w_d$ crosses the
phantom divide line.

In the Chaplygin gas model, at $t=t_0$ (where $w_d(t_0)=-1$),
using (\ref{3}) and (\ref{2}) we obtain
\begin{equation}\label{13}
\dot{\rho_d}(t_0)=-Q(t_0),\,\,\,\rho_d(t_0)=\sqrt{A}.
\end{equation}
Note that nowadays (i.e. $t\simeq t_0$) the dominate part of the
universe is composed of dark sectors. In addition based on
astrophysical data \cite{coin} : ${\rho_{d}(t_0)\over
\rho_{m}(t_0)}={7\over 3}$, therefore the second equation in
(\ref{3}) for a real Hubble parameter requires $\rho_d(t_0)>0$.

The time derivative of the EoS parameter, (\ref{2}), at $t=t_0$
reduces to
\begin{equation}\label{14}
\dot{w_d}(t_0)=-2A\rho_d^{-3}(t_0)Q(t_0)=-2Q(t_0)\rho_d^{-1}(t_0).
\end{equation}
If $Q(t_0)=0$ then $\dot{w_d}(t_0)=0$. In this case, from
(\ref{3}), we deduce $\dot{\rho_d}(t_0)=0$ and to obtain the first
nonzero time derivative of $w_d$ at $t=t_0$ we must continue this
procedure: By taking another time derivative from (\ref{2}), and
then using $\ddot{\rho_d}(t_0)=-\dot{Q}(t_0)$, which follows from
(\ref{3}) when $w_d=-1$, $\dot{w_d}=-1$ and $\dot{\rho_d}=0$, we
obtain
\begin{equation}\label{a}
\ddot{w_d}(t_0)=-2\rho_d^{-1}(t_0)\dot{Q}(t_0).
\end{equation}
If again $\dot{Q}(t_0)=0$ we get $\ddot{w_d}(t_0)=0$. For a
differentiable Hubble parameter, this method can be continued to
give $\rho_d^{(n)}=-Q^{(n-1)}$,
$w_d^{(n)}(t_0)=2A\rho_d^{-3}(t_0)\rho_d^{(n)}(t_0)$, where
$(n-1)$ is the order of the first non zero derivative of $w_d$ at
$t_0$. Therefore finally we obtain
\begin{equation}\label{b}
w_d^{(n)}(t_0)=-2\rho_d^{-1}(t_0)Q^{(n-1)}(t_0).
\end{equation}

If the dark energy sector reaches (or tends to) $w_d=-1$ from the
quintessence phase, depending on whether it will enter in phantom
phase or not, one of the following predicates must be true (we
assume that at $t_0$, $w_d$ is differentiable):

i) $w_d$ tends asymptotically to $-1$. In this case, following
(\ref{b}), all the time derivatives of $w_d$ vanish at $w_d=-1$,
hence $Q$ and all of his time derivatives must also vanish in this
limit:
\begin{equation}\label{15}
Q(t_0)=0,\,\,\,{d^n Q\over
dt^n}(t_0)=:Q^{(n)}(t_0)=0,\,\,\,\forall n\in N.
\end{equation}
For a derivable continuous $w_d$ this can only occur at $t_0\to
\infty$. E.g. the noninteracting Chaplygin gas model, $Q=0$,
belongs to this category and as we have seen $w_d=-1$ occurs
asymptotically. As another example consider
\begin{equation}\label{16}
Q=\lambda(P_d+\rho_d)=\lambda(w_d+1)\rho_d,
\end{equation}
where $\lambda$ is a constant. This is the interaction considered
for some scalar field ($\phi(t)$) models of dark energy or
inflaton, giving rise to the interaction $\lambda \dot{\phi}^2$
\cite{kolb}. It is clear that at $w_d=-1$, we have $Q=0$. Using
(\ref{3}) one can show that the higher derivatives of $Q$ also
vanish at $w_d=-1$, therefore $w_d=-1$ can only occur
asymptotically.

ii) The first nonzero derivative of $w_d$ is negative and of odd
order at $w_d=-1$, in this case after reaching, $w_d$ crosses
$w_d=-1$ line and the Chaplygin gas enters the phantom phase. In
this situation, $Q(t_0)>0$ or, if $Q(t_0)=0$ at $t_0$ the first
nonzero derivative of $Q$ is positive and of even order
\begin{equation}\label{17}
Q(t_0)>0,\,\,\, or \{Q(t_0)=0,\,\,\, and
\,\,\,Q^{(2n)}(t_0)>0,\,\,\,\,\,Q^{(k<2n)}(t_0)=0\}.
\end{equation}

The example proposed in the previous section (\ref{8}), belongs to
this category. The condition $\Gamma>0$ guarantees $Q>0$.

 iii) $\dot{w_d}=0$ at $w_d=-1$, and the first non-vanishing
derivative of $w_d$ is positive and of even order . In this case
the dark component does not enter in the phantom phase and
$w_d=-1$ is the global minimum of $w_d$. In this situation, with
the same method used in ii), it is easy to show that at $t_0$ we
have $Q=0$ and the first nonzero derivative of $Q$ is negative and
of odd order
\begin{equation}\label{18}
Q(t_0)=0,\,\,\,Q^{(2n+1)}(t_0)<0,\,\,\,\,Q^{(k<2n+1)}(t_0)=0.
\end{equation}
In this case the phantom phase is neither accessible for the dark
energy component nor for the universe: $w>-1$.

Therefore in the context of the Chaplygin gas model, reaching or
crossing $w_d=-1$ may not possible for an arbitrary $Q$, e. g.
following (i) and (ii) and (iii) models with $Q< 0$ do not arrive
at $w_d=-1$. Besides, via considering $Q$ as a function of
$\rho_m$ and $\rho_d$, the above conditions give some relationship
between the density of dark matter to density of dark energy at
$t_0$. In this view one can determine or put some conditions on
the value of $r={\rho_m\over \rho_d}$ at the present epoch where
based on some astrophysical data $w_d\approx -1$ is happened. So
there may be a deep relationship between coincidence problem and
$w_d=-1$ crossing observed in the present era.

To illustrate this point, let us consider the interaction
\cite{examp}
\begin{equation}\label{19}
Q=H(\lambda_m\rho_m+\lambda_d\rho_d),
\end{equation}
and study the conditions required to reach $w_d=-1$. Following (i)
and (ii) and (iii) we must have $Q\geq 0$ at $t_0$. For an
expanding universe this implies $ \lambda_m r(t_0)\geq
-\lambda_d$.

For
\begin{equation}\label{1000}
r(t_0)= -{\lambda_d\over \lambda_m},
\end{equation}
$Q=0$ and $\dot{w_d}=0$, but using $\dot{\rho_d}(t_0)=0$ and
$\dot{\rho_m}(t_0)=-3H\rho_m(t_0)$ we obtain
$\ddot{w_d}(t_0)=-3\lambda_d H^2(t_0)$ and following (ii) (in
order that reaching $w_d=-1$ be allowed) we must have
$\lambda_d<0$. In this situation the dark energy component,
remains in the quintessence phase. This possibility is excluded
when one considers the interaction $Q=\lambda H(\rho_m+\rho_d)$
which is adopted in some papers \cite{inter1}.

If $\lambda_m r> -\lambda_d$ at $t_0$, then $w_{d0}$ cross the
phantom divide line. In this case for $\lambda_m>0,\,\,\,$ and
$\lambda_d<0$ we obtain a lower bound for $r(t_0)$:
\begin{equation}\label{1001}
r(t_0)> -{\lambda_d\over \lambda_m}
\end{equation}
at transition time.

Although obtaining an exact solution to the Friedmann equations in
the presence of interaction (\ref{19}) does not seem to be
possible but let us examine whether these equations admit the
series solution
\begin{eqnarray}\label{20}
w_d&=&-1+w_{d0}t^\alpha+\mathcal{O}(t^{\alpha+1}),\nonumber \\
\rho_d&=&\rho_{d0}+\rho_{d1}t^\beta+\mathcal{O}(t^{\beta+1}),\nonumber
\\
\rho_m&=&\rho_{m0}+\rho_{m1}t^\gamma+\mathcal{O}(t^{\gamma+1}),\nonumber\\
H&=&H_0+H_1t+H_2t^2+\mathcal{O}(t^3),
\end{eqnarray}
at $t=0$ (we have taken $t_0=0$). (\ref{1}) leads to
$\alpha=\beta$,\,\,\, $\rho_{d0}=\sqrt{A}$ and
$\rho_{d1}={w_{d0}\sqrt{A}\over 2}$.  We have also assumed that
$\dot{H}(t_0)\neq 0$ which lies on the assumption
$\Omega_d(t_0)\neq 0$. We first consider the solution with
$\alpha\geq 2$. Putting (\ref{20}) into the first equation in
(\ref{3}) and equating the terms with the same power of $t$, after
some calculation we obtain
$\lambda_m\rho_{m0}+\lambda_d\rho_{d0}=0$, and $\alpha-1=\gamma$
and subsequently by using the second equation in (\ref{3}) we
obtain $\alpha=2$ (viz. $\alpha>2$ is not allowed),
$\gamma=1$,\,\,\, $\rho_{m1}+3H_0\rho_{m0}=0$ and
\,\,\,$2\rho_{d1}=-\lambda_{m}\rho_{m1}H_0$. By collecting these
together, the series solution for $\alpha\geq 2$ at $t=0$ (which
is also consistent with (\ref{4})) is obtained as
\begin{eqnarray}\label{21}
w_d&=&-1-3\lambda_dH_0^2t^2+\mathcal{O}(t^3)\nonumber \\
\rho_d&=&\sqrt{A}-{3\over
2}\lambda_dH_0^2\sqrt{A}t^2+\mathcal{O}(t^3)\nonumber \\
\rho_m&=&-{\lambda_d\over
\lambda_m}\sqrt{A}(1-3H_0t)+\mathcal{O}(t^2)\nonumber \\
H&=&H_0+4\pi{\lambda_d\over \lambda_m}\sqrt{A}t+\mathcal{O}(t^2),
\end{eqnarray}
where the Hubble parameter at $t=0$ is determined to be
$H_0={8\pi\over 3}\left(1-{\lambda_d\over
\lambda_m}\right)\sqrt{A}$. This solution corresponds to the
system characterized by (\ref{1000}).

Note that $\rho_{m1}+3H_0\rho_{m0}=0$ in this example leads to
$Q(t_0)=0$. This lies on the fact that $\alpha=2$ in (\ref{20}),
and therefore the Chaplygin gas remains in quintessence phase:
$\dot{w_d}(t_0)=0$. Hence (\ref{14}) results in
$Q(t_0)=\dot{w_d}(t_0)=0$. As we will show this is not true for
other possible solutions.

For $\alpha=1$ (corresponding to the aforementioned case
$\lambda_m r(t_0)> -\lambda_d$), one can similarly show that the
following solution exists:
\begin{eqnarray}\label{22}
w_d&=&-1+{2\over
\sqrt{A}}\left((\lambda_m-\lambda_d)\sqrt{A}-{3H_0^2\over
{8\pi}}\lambda_m
\right)H_0t+\mathcal{O}(t^2)\nonumber \\
\rho_d&=&\sqrt{A}+\left((\lambda_m-\lambda_d)\sqrt{A}-{3H_0^2\over
{8\pi}}\lambda_m\right)H_0t+
\mathcal{O}(t^2)\nonumber \\
\rho_m&=&\left({3H_0^2\over {8\pi}}-\sqrt{A}\right)+
\left((\lambda_m-\lambda_d+3)\sqrt{A}-{3H_0^2\over
8\pi}(\lambda_m+3)\right)H_0t+ \mathcal{O}(t^2)
\nonumber \\
H&=&H_0+\left(-{3H_0^2\over
2}+4\pi\sqrt{A}\right)t+\mathcal{O}(t^2).
\end{eqnarray}
According to the condition (ii), (\ref{17}) and subsequently
(\ref{22}) are valid when
\begin{equation}\label{23}
\lambda_m H_0^2>{8\pi\over 3}(\lambda_m-\lambda_d)\sqrt{A}.
\end{equation}
This solution crosses $w_d=-1$. If $\lambda_m=0$ we must have
$\lambda_d>0$ which is the same result discussed in the previous
section. If $\lambda_d=0$, only models with $\lambda_m>0$ are
allowed. If we adopt $\lambda_m=\lambda_d=\lambda$, then $w_d=-1$
crossing is possible whenever $\lambda>0$.

After the dark energy component of the universe passes $w_d=-1$,
the EoS parameter of the universe may cross $w=-1$ too. Note that
based on the values estimated for $w_d$ and $\Omega_d$, it is also
possible that the universe has an EoS parameter satisfying
$w=w_d\Omega_d \simeq -1$ even in the present era, e.g. if we
adopt $ \Omega_d=0.73$ and  $w_d\ \simeq -1.33$ \cite{Komat}, we
obtain $w=-0.97$.

For a general dark energy model it can be shown that
\begin{equation}\label{24}
w=-{1\over 3H}{\dot{\Omega_d}\over {1-\Omega_d}}-{Q\over
{3H\rho_c(1-\Omega_d)}}.
\end{equation}
In the Chaplygin gas, from $w_d\Omega_d=w$ we derive
\begin{equation}\label{25}
w=-A\left({8\pi \over 3}\right)^2{1\over H^4\Omega_d}.
\end{equation}
(\ref{24}) together with (\ref{25}) form a system (which is not
analytically solvable) describing the behavior of $w$ and
$\Omega_d$ in terms of the scale factor. If the system arrives at
$w=-1$ at $t=\tau$ from (\ref{25}) we obtain
\begin{equation}\label{26}
\dot{w}(\tau)={\dot{\Omega_d}(\tau)\over \Omega_d(\tau)}.
\end{equation}
The universe enters in the phantom phase provided that
$\dot{w}(\tau)\leq 0$. By considering (\ref{24}), this implies
\begin{equation}\label{27}
Q\geq 3H\rho_c(1-\Omega_d),
\end{equation}
at $t=\tau$. In terms of $\Omega_m={\rho_m\over \rho_c}$ this
inequality becomes
\begin{equation}\label{28}
Q\geq 3H\rho_m
\end{equation}
at $t=\tau$. Putting (\ref{19}) in this inequality gives
\begin{equation}\label{29}
(\lambda_m-3)r\geq -\lambda_d,
\end{equation}
which for $\lambda_m>3$ and $\lambda_d<0$, gives a lower bound for
$r$ at transition time:
\begin{equation}
r\geq -{\lambda_d\over{\lambda_m-3}},
\end{equation}
while for $\lambda_m<3$ and $\lambda_d>0$,
$-{\lambda_d\over{\lambda_m-3}}$ becomes an upper bound for $r$.
For $\lambda=\lambda_d=\lambda_m$, the upper bound becomes
$r<{\lambda\over 3-\lambda}$.

To study the behavior of the system near $w=-1$ as before we may
consider the following series solutions for $H$ and $\Omega_d$ at
$t=\tau$ (note that $\dot{H}(\tau)=0$),
\begin{eqnarray}\label{30}
H&=&h_0+h_1(t-\tau)^\nu+\mathcal{O}((t-\tau)^{\nu +1}),\,\,\nu\geq 2,\,\,\, h_1>0 \nonumber \\
\Omega_d&=&u_0+u_1(t-\tau)^\mu+\mathcal{O}((t-\tau)^{\mu+1}).
\end{eqnarray}
$\mu$ and $\nu$ are the orders of the first nonzero derivatives of
$H$ and $\Omega_d$ at $t=\tau$ respectively. As we are studying
the transition from quintessence to phantom phase we have taken
$h_1>0$. Using (\ref{24}) and the equation
\begin{equation}\label{31}
w=-1-{2\over 3}{\dot{H}\over H^2},
\end{equation}
we find that if $\mu\neq 1$ then we must have $\mu=\nu$ (for
details see \cite{sad2}), but inserting the series solution in
(\ref{24}) yields $\nu=\mu+1$ which results in $\nu=2$, and
$\mu=1$. Note that $\nu=2$ implies that the system cannot remain
in quintessence phase. We also obtain
$u_1=\left(3-\lambda_m\right)h_0+\left(\lambda_m-\lambda_d-3\right)h_0u_0$.
Using (\ref{26}), (\ref{3}), and (\ref{4}) we arrive at
\begin{eqnarray}\label{32}
H&=&h_0-{3\over 4}h_0^3\left({9\over 64\pi^2 A}(3-\lambda_m)h_0^4+
(\lambda_m-\lambda_d-3)\right)(t-\tau)^2\nonumber \\
&&+\mathcal{O}((t-\tau)^{3})\nonumber \\
\Omega_d&=&{64\pi^2A\over{9h_0^4}}+\left((3-\lambda_m)h_0+{64\pi^2\over
9h_0^3}(\lambda_m-\lambda_d-3)\right)(t-\tau)\nonumber\\
&&+\mathcal{O}((t-\tau)^{2}).
\end{eqnarray}
$h_0$ is not determined, but $h_1>0$ gives
\begin{equation}\label{33}
(3-\lambda_m)h_0^4< {64\pi^2 A\over 9}(3-\lambda_m+\lambda_d).
\end{equation}
This inequality, which using $r={1-\Omega_d\over \Omega_d}$ can be
shown to be the same as (\ref{29}), determines the relation
between the energy density at transition time and the parameters
of the model.

\section{Conclusion}
We considered the interacting Chaplygin gas as the dark energy
component of the universe. Some necessary conditions on the
interaction, allowing the Chaplygin gas to reach the phantom
divide line ($w_d=-1$), were discussed.  Using these condition we
found that the ratio of dark matter to dark energy density, $r$,
must satisfy some inequalities (or equalities) which can alleviate
the coincidence problem, i.e. at $w_d=-1$, $r$ is whether
determined in terms of the parameters of the system or a nonzero
lower bound can be found for it. Via some examples, we showed that
how the form of interaction determines some of the behavior of the
system such as crossing the phantom divide line or remaining in
the quintessence phase. We also obtained a series solution to
Friedmann equations at phantom divide line. At the end via an
example we studied the possibility that the EoS of the universe
crosses $w=-1$.

\end{document}